\documentclass[preprint,aps]{revtex4}
\usepackage{graphicx}
\usepackage{psfrag}
\newcommand{\lsim}{\mathrel{\mathop{\kern 0pt \rlap
  {\raise.2ex\hbox{$<$}}}
  \lower.9ex\hbox{\kern-.190em $\sim$}}}
\newcommand{\gsim}{\mathrel{\mathop{\kern 0pt \rlap
  {\raise.2ex\hbox{$>$}}}
  \lower.9ex\hbox{\kern-.190em $\sim$}}}
\newcommand{\beq}     {\begin{equation}}
\newcommand{\eeq}     {\end{equation}}
\newcommand{\bea}     {\begin{eqnarray}}
\newcommand{\eea}     {\end{eqnarray}}

\newcommand{\dt}      {\delta}

\newcommand{\lm}      {\lambda}
\newcommand{\gm}      {\gamma}
\newcommand{\Gm}      {\Gamma}
\newcommand{\sg}      {\sigma}
\newcommand{\kp}      {\kappa}
\newcommand{\rd}      {\partial}
\newcommand{\Lg}       {{\mathcal L}}
\newcommand{\M}       {{\mathcal M}}
\newcommand{\n}       {{\vec{n}}}
\newcommand{\no}      {\nonumber}
\newcommand{\phono}      {{\tilde{\gamma}}}
\newcommand{\Gno}      {{\tilde{G}}}
\newcommand{\Lmsusy}      {\Lambda_{_{\rm SUSY}}}
\newcommand{\Mpl}      {M_{{\rm Pl}}}

\begin{document}
\preprint{KIAS02030}

\title{
Kaluza-Klein gravitino production \\
with a single photon at $e^+e^-$ colliders}

\author{Seungwon Baek,~~ Seong Chan Park~~ and~~
Jeonghyeon Song}
\affiliation{%
Korea Institute for Advanced Study\\
207-43 Cheongryangri-dong, Dongdaemun-gu, Seoul 130-012, Korea}%

\date{\today}

\vspace{3cm}
\begin{abstract}
In a supersymmetric large extra dimension scenario,
the production of Kaluza-Klein gravitinos accompanied by a photino
at $e^+ e^-$ colliders is studied.
We assume that a bulk supersymmetry
is softly broken on our brane such that the low-energy theory
resembles the MSSM.
Low energy supersymmetry breaking is further assumed as in GMSB,
leading to
sub-eV mass shift in each KK mode of the gravitino
from the corresponding graviton KK mode.
Since the photino decays within a detector due to
the sufficiently large inclusive decay rate of $\phono\to\gm\,\Gno$,
the process $e^+ e^- \to \phono\,\Gno$ yields
single photon events with missing energy.
Even if the total cross section can be substantial at
$\sqrt{s}=500$ GeV,
the KK graviton background of $e^+ e^- \to \gm G$ is
kinematically advantageous and thus much larger.
It is shown that the observable
$\Delta \sg_{LR} \equiv \sg(e^-_L e^+_R)-\sg(e^-_R e^+_L)$ can
completely eliminate the KK graviton background but
retain most of the KK gravitino signal,
which provides a unique and robust method
to probe the \emph{supersymmetric} bulk.

\end{abstract}

\maketitle

\section{Introduction}
\label{introduction}
The standard model (SM) has been thoroughly tested in various experiments
even if the Higgs boson remains as the only missing ingredient.
In the theoretical view point, however,
the SM has several unsatisfactory aspects such as the gauge hierarchy problem:
The Higgs boson mass near the electroweak scale requires a fine tuning
to eliminate quadratically divergent radiative corrections.
Low-energy supersymmetry is known to cancel the quadratic divergence
by introducing a supersymmetric partner for each SM particle \cite{SUSY}.
Supersymmetry protects the electroweak scale Higgs mass
from the Planck scale,
as the chiral (gauge) symmetry does for fermions (gauge bosons).

Recently another new route to the solution of the gauge hierarchy
problem has been opened based on the advances in string theories, by
introducing extra dimensions.
Arkani-Hamed, Dimopoulos and
Dvali (ADD) proposed that the large volume of $\dt$-dimensional extra dimensions
can explain the observed huge Planck scale $M_{\rm Pl}$ {\cite{Antoniadis:1998ig}}:
The fundamental gravitational scale or string scale $M_S$ is related with the
Planck scale $M_{\rm Pl}$ and the size of an extra dimension $R$
by $\Mpl^2=M_S^{\dt+2} R^\dt$;
the hierarchy problem is
resolved as $M_S$ can be maintained around TeV.
Later
Randall and Sundrum (RS) proposed another higher dimensional
scenario based on two branes and a single extra dimension
compactified in a slice of anti-deSitter space {\cite{Randall:1999ee}}:
The hierarchy problem is explained by
a geometrical exponential factor.
Very interesting is that these extra dimensional models
can lead to distinct and rich phenomenological signatures
in the future colliders,
characterized by low-energy gravity effects \cite{Wells,ED-ph}.
In ADD case, for example, the multiplicity of gravitons
below an energy scale $E$ is proportional to $(E R)^\dt(=\Mpl^2 E^\dt/M_S^{\dt+2})$,
which is extremely large and compensates the small
gravitational coupling.

An economical description of new physics
to solve the gauge hierarchy problem
would introduce either low-energy supersymmetry or extra dimensions.
Nevertheless supersymmetric
bulk is theoretically more plausible~\cite{susyed,Altendorfer:2000rr} since
the most realistic framework of extra dimensional models,
string/M theory\cite{MTheory},
indeed possesses supersymmetry as a fundamental symmetry.
Moreover, extra dimensions can play the role of supersymmetry breaking
on a
hidden brane~\cite{arkani1} or
in the bulk
by Scherk-Schwarz compactification~\cite{antoniadis}.

Obviously phenomenological signatures of supersymmetric
bulk are crucially dependent on how many supersymmetries
survive on our brane below the scale $M_S$.
One interesting possibility is that
a single supersymmetry is softly broken on our brane
such that our low-energy effective theory
yields supersymmetric spectra as in the
minimal supersymmetric standard model (MSSM).
One distinctive feature of this scenario
is the presence of the gravitino, the superpartner of the graviton.
Since this gravitino also propagates in the full dimensional space
as it belongs to the same
supermultiplet with the graviton,
we have gravitino Kaluza-Klein modes on our brane.
The soft and spontaneous breaking of a supersymmetry results in
the mass shift between a graviton KK mode
and the corresponding gravitino KK mode,
of order $\Lmsusy^2/\Mpl$.
Here the four-dimensional Planck mass $\Mpl$
scales the strength of gravitino coupling
and $\Lmsusy$ is the supersymmetry breaking scale.
If low-energy supersymmetry breaking is assumed,
e.g., in the gauge mediated supersymmetry breaking (GMSB) scenario,
the resulting mass shift is very light:
For $\Lmsusy\sim 100$ TeV,
$\Lambda_{_{\rm SUSY}}^2/\Mpl \sim 1$ eV.
Restricting ourselves to the ADD scenario,
we have almost continuous spectrum of KK gravitinos
with the zero mode mass at sub-eV scale.
In Ref.~\cite{Hewett:2002uq},
the four-dimensional effective theory
in a supersymmetric ADD scenario
has been derived, including the couplings of the bulk gravitino KK states
to a fermion and its superpartner.
At $e^+ e^-$ colliders,
the virtual exchange of KK gravitinos
can occur only
in the selectron pair production
which was shown to
substantially enhance the total cross section
and change the kinematic distributions.

Another distinctive signature of KK gravitinos
is their production at high-energy colliders.
A superlight gravitino, which becomes the stable lightest
supersymmetric particle (LSP), escapes any detector,
leading to missing energy events.
Moreover
the decay modes of a supersymmetric particle $\widetilde{X}$ are now changed.
Even if the $\widetilde{X}$ is the lightest
among supersymmetric partners of the SM particles, e.g., the photino,
a new decay mode of $\phono \to \gm \,\Gno$ is opened and dominant.
As shall be discussed later,
this decay rate is large enough for the photino
to decay within a detector.
Therefore,
the process $e^+ e^- \to \phono\,\Gno$ yields
a typical signature of a single photon at large transverse momentum.
And the summation over
all possible extra-dimensional momenta yields a sizable
production rate
characterized by the $M_S$ scale.
This process has kinematic advantages over the selectron pair production
in case the selectron is
too heavy to be pair-produced.

Of great significance is to signal not only the extra dimensions,
but also the \emph{supersymmetric}
extra dimensions, i.e., KK gravitinos.
Unfortunately, single photon events with
missing energy in this scenario have two more sources,
the SM process of $e^+ e^- \to \gm \nu\overline{\nu}$ and
the KK graviton production of $e^+ e^- \to \gm G$.
With an appropriate cut to reduce the $Z$-pole contributions
of the SM,
the KK graviton production can be compatible
with the SM background at the future $e^+ e^-$ collider.
However the KK gravitino production rate
is smaller than the KK graviton case by an order of magnitude.
This is due to the kinematic suppression by the production of the massive photino
while the dependence of the $M_S$ and $\dt$ is the same
for both the KK graviton and gravitino production.
Total cross section alone cannot tell whether the bulk
possesses supersymmetry or not.
We shall show that the observable
$\Delta \sg_{LR} \equiv \sg(e^-_L e^+_R)-\sg(e^-_R e^+_L)$
completely eliminates the KK graviton effects, but retains
most of the KK gravitino effects.
This is because in a supersymmetric model
the sign  of the coupling of
a left-handed electron with a photino (and a selectron)
is opposite to that of a right-handed electron (a left-handed $anti$-electron).
The coupling with gravitino, which is gravitational,
does not depend on the fermion chirality.
Therefore, the scattering amplitudes of
the $t$- and $u$-channel diagrams,
where both couplings are involved, have opposite
sign for $e^-_L$ and $e^-_R$ beams.
As the $t$- and $u$-channel amplitudes
are added to
the $s$-channel one, mediated by a photon,
the total scattering amplitudes are different for the left- and right-handed
electron beam.
For the KK graviton production accompanied by a single photon,
all the involved interactions are chirality blind
so that the corresponding $\Delta \sg_{LR}$ vanishes.

Our paper is organized as follows.
In Sec.~\ref{effLg}, we review the four-dimensional effective Lagrangian
in a supersymmetric ADD scenario.
And analytic expressions for photino decay rate into a photon and KK gravitinos
and for the process $e^+ e^- \to \phono\,\Gno$ are to be given.
Section \ref{Num} devotes to the phenomenological
discussions of this scenario, including
total cross section, kinematic distributions,
a specific observable by using the polarization of electron beam
and so on.
In Sec.~\ref{Conc} we give our conclusions.

\section{Effective Gravitino Lagrangian}
\label{effLg}
In this paper,
we assume that there are $\dt$ large and supersymmetric extra dimensions,
and a single supersymmetry is softly broken on our brane
such that our low-energy effective theory
yields the
MSSM spectra.
The cases of more than three extra dimensions are
to be considered
since in the $\dt=2$ case
astrophysical and cosmological constraints are
too strong that
the $M_S$ is pushed up to about 100 TeV,
disfavored as a solution of the gauge hierarchy problem \cite{astro}.
And the MSSM super-particles are assumed to be confined on our brane.
New feature is then
another KK tower of the gravitino.
The compactification of the gravitino field
in a supergravity theory leads to the four-dimensional effective action
which is a sum of KK states of massive spin 3/2 gravitinos~\cite{Hewett:2002uq}.
The free part of the effective Lagrangian gives the propagator of
the $\vec{n}$-the KK mode of the gravitino with momentum $k$ and mass $m_n$ such as
\begin{equation}
\frac{i {\mathcal P}^\n_{\mu\nu}}{k^2-m_\n^2}.
\end{equation}
Here ${\mathcal P}^\n_{\mu\nu}$ is
\begin{eqnarray}\label{P}
{\mathcal P}^{\vec{n}}_{\mu\nu}
&\equiv&
\sum_{\lm}\Gno_\mu^{\vec{n}}(k,\lm)\overline{\Gno^{\vec{n}}_\nu}(k,\lm)
\\ \no
&=&(\rlap{\hspace{.5mm}/}{k}+m_n) \left( \frac{k_\mu
k_\nu}{m_n^2}-\eta_{\mu\nu}\right) -\frac{1}{3} \left(
\gm^\mu+\frac{k^\mu}{m_n} \right) (\rlap{\hspace{.5mm}/}{k}-m_n)
\left( \gm^\nu+\frac{k^\nu}{m_n} \right),
\end{eqnarray}
where $\gm^\mu{\mathcal P}^{\vec{n}}_{\mu\nu}=0$ and
$k^\mu{\mathcal P}^{\vec{n}}_{\mu\nu}=0$.

The effective interaction Lagrangian for the KK gravitino is obtained
from the general Noether technique,
irrespective to the detailed supersymmetry breaking mechanism.
The coupling of each KK mode of graviton and gravitino
is determined by the Planck constant
\begin{equation}\label{kappa}
   \Mpl^{-1} \equiv \kp=\sqrt{8\pi G_{\rm Newton}}
   \approx\frac{1}{2.4\times 10^{18}~{\rm GeV}}.
\end{equation}
Minimally coupled to gravity,
the interactions of a KK gravitino with a fermion and photon field
to leading order in $\kp$ are \cite{Mendez:1984qs}
\begin{eqnarray}
\label{LgffG}
\Lg_{f \tilde{f} \Gno} &=&
-\frac{\kp}{\sqrt{2}}\left[
\overline{\Gno}_\mu \gm^\nu \gm^\mu \psi_L \rd_\nu\phi_L^*
+\overline{\Gno}_\mu \gm^\nu \gm^\mu \psi_R \rd_\nu\phi_R^*
+h.c.
\right], \\
\label{LgrrG}
\Lg_{\gm\phono\Gno}
&=&
-\frac{\kp}{4}\;\overline{\phono}\,\gm^\mu
[\gm^\rho,\gm^\sg]\Gno_\mu\rd_\rho A_\sg
+\frac{\kp}{4}\;\overline{\Gno}_\mu
[\gm^\rho,\gm^\sg]\,\gm^\mu\,\phono\;\rd_\rho A_\sg
\,.
\end{eqnarray}
For later discussion, we present the interaction Lagrangian
for the electron-selectron-photino:
\begin{equation}
\label{Lgffr}
\Lg_{f \widetilde{f}\;\phono}
=
-\sqrt{2}\,e\, Q_f \left[
\overline{\phono}_R \psi_L \phi^*_L +\overline{\psi}_L \phono_R\phi_L
-\overline{\phono}_L \psi_R \phi^*_R -\overline{\psi}_R \phono_L\phi_R
\right]
\,.
\end{equation}

Since each KK mode of gravitons and gravitinos escapes a detector,
experimentally applicable are
inclusive rates with all the kinematically allowed KK modes
summed up.
Due to the very small mass splitting among KK modes,
the summation can be
approximated by a continuous integration over the KK mode mass $m$
such as  \cite{Wells}
\begin{equation}\label{sum}
    \sum_{\vec{n}} \rightarrow
    \int d m \frac{\Mpl^2 m^{\dt-1}}{ M_S^{2+\dt}}S_{\dt-1}\,,
\end{equation}
where $S_{\dt-1}$ is the volume of the unit sphere in
$\dt$ dimensions, given by $S_{\dt-1}=2\pi^{\frac{\dt}{2}}/\Gamma(\dt/2)$.
The $\Mpl^2$ in the numerator,
implying the tremendous number of accessible KK modes,
compensates the gravitational coupling.
In effect, the Planck scale is lowered
to the $M_S$ of TeV scale.

\subsection{Decay rate of a photino}

It has been known that the presence of a light gravitino
alters the decay modes of supersymmetric particles
as the gravitino becomes the LSP;
the decay mode of $\widetilde{X} \to X\Gno$ becomes dominant
\cite{Mendez:1984qs,GmGMSB}.
Even though the coupling strength of the gravitino
is Planck-suppressed,
the wave function of a light gravitino with mass
$m_{3/2}$,
momentum $k^\mu$ and helicity $\pm 1/2$
is an ordinary spin 1/2 wave function multiplied by
the large factor $\sqrt{2/3}\,k^\mu/m_{3/2}$ \cite{Mendez:1984qs}.
The gravitino mass $m_{3/2}$, e.g.,
in the gauge mediation supersymmetry breaking (GMSB)
where the supersymmetry breaking scale is generically low,
is
\begin{equation}\label{mGno}
   m_{3/2}=\frac{\Lambda^2_{_{\rm SUSY}}}{\sqrt{3}\Mpl} \simeq
   2.36 \left(\frac{\Lmsusy}{100~{\rm TeV}}
   \right)^2 {\rm eV}.
\end{equation}
Thus the $\Mpl$ term in the $m_{3/2}$ cancels
the gravitational coupling $\Mpl$, so that
the characteristic scale of the decay rate
becomes the supersymmetry breaking scale $\Lmsusy$.
The photino decay rate is known to be \cite{GmGMSB}
\begin{equation}
\label{GmGMSB}
\Gamma(\phono \to \gm \Gno)
=
\frac{1}{48\pi}
\frac{M_\phono^5}{\Mpl^2 m_{3/2}^2}
= \frac{1}{16\pi} \frac{M_{\phono}^5}{\Lambda^4_{_{\rm SUSY}}}
,
\end{equation}
where $M_\phono$ is the photino mass.
For $\Lmsusy \lsim 10^3$ TeV with $M_\phono\sim 100$ GeV,
the photino decays within a CDF-type detector.

In a supersymmetric ADD scenario,
a photino can decay into a photon and a KK mode of a gravitino,
if kinematically allowed.
The decay rate for the $\n$-th KK gravitino is
\begin{equation}\label{Gmm}
    \Gamma(\phono \rightarrow \gm \,\Gno^\n)=
    \frac{1}{48 \pi}
    \frac{\kp^2 M_{\phono}^5 }{m_{n}^2}
    \left(
    1-\frac{m_n^2}{M_\phono^2}
    \right)^3
    \left(
    1+3\frac{m_n^2}{M_\phono^2}
    \right).
\end{equation}
The inclusive decay rate of a photino
is obtained by the sum in Eq.~(\ref{sum}):
\begin{equation}\label{Gm}
\Gm_{\rm tot}
\equiv\sum_\n\Gm(\phono \rightarrow \gm \Gno^\n)=
\frac{f_\dt}{16\pi}\frac{M_\phono^5 }{M_S^4}
  \left(\frac{M_\phono}{M_S}
 \right)^{\dt-2},
\end{equation}
where $f_\dt \equiv 64 \, S_{\dt-1}/\{(\dt^2-4)(\dt+4)(\dt+6)\}$ of order one.
Numerically $f_3 \approx 2.55$, $f_4 \approx 1.32$,
$f_5 \approx 0.81$, and $f_6 \approx 0.52$.
Here one should note that
the $\Gm_{\rm tot}$ does not depend on
the exact value of $\Lmsusy$ which determines
the zero mode mass of the KK gravitino,
as long as the supersymmetry breaking ensures a superlight gravitino.

In general, the decay rate $\Gm_{\rm tot}$ is quite large for $M_S$
of order TeV even with the suppression of
$({M_\phono}/{M_S}
 )^{\dt-2}$.
For various number of extra dimensions,
the magnitude of the inclusive photino decay rate is
\begin{equation}
\label{GmN}
\Gm_{\rm tot}
= \left(
\frac{M_\phono}{100~{\rm GeV}}
\right)^{\dt+3}
\left(
\frac{1~{\rm TeV}}{M_S}
\right)^{\dt+2} \times
\left\{\begin{array}{c}
  50.8~{\rm keV}\quad {\rm for}\quad \dt=3 \\
  2.62~{\rm keV}  \quad {\rm for}\quad \dt=4 \\
  0.16~{\rm keV}  \quad {\rm for}\quad \dt=5 \\
  0.01~{\rm keV}  \quad {\rm for}\quad \dt=6 \\
\end{array}
 \right.
\,.
\end{equation}
Then, the average distance travelled by an photino with energy $E$ in
the laboratory frame is
\begin{equation}
\label{L}
L
= \left({E^2}/{M_\phono^2}-1\right)^{\frac{1}{2}}\left(
\frac{100~{\rm GeV}}{M_\phono}
\right)^{\dt+3}
\left(
\frac{M_S}{1~{\rm TeV}}
\right)^{\dt+2} \times
\left\{\begin{array}{c}
  4.0 \times 10^{-10}~{\rm cm}~~ {\rm for}\quad \dt=3 \\
  7.7 \times 10^{-9}~{\rm cm}  \quad {\rm for}\quad \dt=4 \\
  1.3 \times 10^{-7}~{\rm cm}  \quad {\rm for}\quad \dt=5 \\
  1.8 \times 10^{-6}~{\rm cm}  \quad {\rm for}\quad \dt=6 \\
\end{array}
\right.
\,.
\end{equation}
Thus the photino decays within a detector, leaving a detectable photon signal.
In the following,
we investigate at $e^+ e^-$  collisions
the production of a KK gravitino and a photino,
which generates single photon events with missing energy.

\subsection{Cross Section of $e^+e^- \to \phono \Gno$}

\begin{figure}
\includegraphics{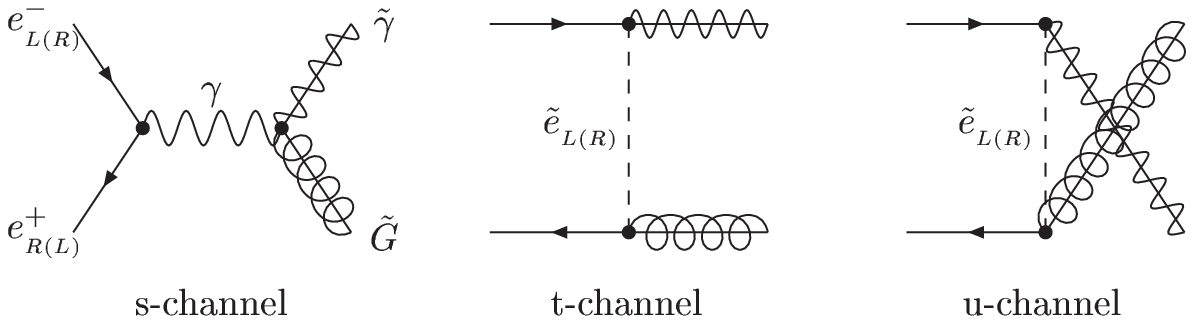}
\caption{\label{feyn} Feynman diagrams for
the process $e^+ e^- \to \phono\Gno$.}
\end{figure}

For the process
\begin{equation}
    e^-(p_1,\lm_e)+e^+(p_2,\overline{\lm}_e) \longrightarrow
\phono(k_1)+\Gno^\n(k_2)
,
\end{equation}
there are three Feynman diagrams mediated by the selectron and photon
as depicted in Fig.~\ref{feyn}.
The Mandelstam variables are defined by
$s=(p_1+p_2)^2$, $t=(p_1-k_1)^2$, and $u=(p_1-k_2)^2$.
Then the helicity amplitudes apart from $i \kp e$ factor, defined by
$
\M(\lm_e,\overline{\lm}_e) \equiv i \kp\,{e}
\widehat{\M}^{\lm_e}
$,
are
\begin{eqnarray}\label{Mpol}
\widehat{\M}^\mp
&=&
\bar{v}_e(p_2) \gm^\mu P_\mp u_e(p_1) \\ \no &\times&
\overline{\Gno}_\nu(k_2)
\left[\pm
\frac{1}{t-\widetilde{m}_{e_\mp}^2}(p_1-k_1)^\nu\gm_\mu P_\mp
\mp
\frac{1}{u-\widetilde{m}_{e_\mp}^2}(p_1-k_2)^\nu\gm_\mu P_\pm
\right.
\\ \no
&& \left.\qquad\quad
-\frac{1}{4s}[\rlap{\hspace{.5mm}/}{k}_1+\rlap{\hspace{.5mm}/}{k}_2,\gm_\mu]\gm^\nu
\right]v_\phono(k_1) \,,
\end{eqnarray}
where $P_\pm=(1\pm \gm^5)/2$ and
$\widetilde{m}_{e_{-(+)}}\hspace{-0.2cm}=\widetilde{m}_{e_{L(R)}}$.

The differential cross section is then
\begin{eqnarray}\label{sgtot}
\frac{d^2 \sg}{dx_\phono  d \cos\theta_\phono}
\left( e^+ e^- \to \phono\,\Gno\right)
&=&\frac{\alpha}{32} \, S_{\dt-1}
\left( \frac{\sqrt{s}}{ M_S}\right)^{\dt+2}
\frac{1}{s}\,
\left(
1+\frac{M_\phono^2}{s}-x_\phono
\right)^{\dt/2-1} \\ \no && \times
\sqrt{\lm_\phono}\,
f_\Gno(x_\phono,\cos\theta)
\,,
\end{eqnarray}
where $x_\phono \equiv 2 E_\phono/\sqrt{s}$ and
$\lm_\phono\equiv \lm(1,{M_\phono^2}/{s},
1+{M_\phono^2}/{s}-x_\phono)$.
Here $\lm(a,b,c)=a^2+b^2+c^2-2 a b-2 b c -2 a c$,
 and
\begin{equation}\label{fGno}
f_\Gno(x_\phono,\cos\theta)\equiv
\frac{
\left| \widehat{\M}^-
\right|^2
+
\left| \widehat{\M}^+
\right|^2
}{2\,s}
\,.
\end{equation}
The range of $x_\phono$ is $[2M_\phono/\sqrt{s},1+M_\phono^2/s]$.
The amplitudes squared are summarized in the Appendix.
It is to be compared to the KK graviton production process:
\begin{equation}\label{sigG}
\frac{d^2 \sg}{dx_\gm  d \cos\theta}
\left( e^+ e^- \to \gm\, G\right)
=\frac{\alpha}{32} \, S_{\dt-1}
\left( \frac{\sqrt{s}}{ M_S}\right)^{\dt+2}
\frac{1}{s}\, (1-x_\gm)^{(\dt/2-1)}
f_G(x_\gm,\cos\theta)
\,,
\end{equation}
where $x_\gm\equiv 2E_\gm/\sqrt{s}$ and for $f_G$
we refer to Ref.~\cite{Wells}.

Equations (\ref{sgtot}) and (\ref{sigG})
show that both the differential cross sections
have the same $M_S$-dependence.
The gravitino-production accompanied by a
massive photino is at a kinematically disadvantage,
relative to the graviton-production with a massless photon.
The measurement of total cross section alone is not
enough to probe \emph{supersymmetric} bulk effects.
Some kinematic distributions and other observables are needed.

We notice that there is one crucial characteristic for the
gravitino production accompanied by a photino.
As explicitly shown in Eq.~(\ref{Lgffr}),
the coupling sign of
a left-handed electron with a photino and a selectron
is opposite to that of a right-handed electron:
The holomorphy of the super-potential requires
that a fermion should belong to a (left-handed) chiral superfield;
the right-handed electron is to be described by
a left-handed $anti$-electron, which possesses positive charge.
The interaction with a gravitino, which is gravitational, does not
distinguish the chirality of the involved fermion.
Therefore,
the scattering amplitudes of
the $t$- and $u$-channel diagrams,
which include one $e-\widetilde{e}-\phono$ and one $e-\widetilde{e}-\Gno$ coupling,
have opposite sign for the left- and right-handed electron beam.
In the $s$-channel diagram,
the electron is coupled with the ordinary QED photon.
Since two kinds of amplitudes
(one changes the sign under the helicity flip of the electron beam,
whereas the other does not) are added,
we are ended up with chirality-sensitive total cross section.
Note that without the $s$-channel diagram,
the sign-change in the amplitudes alone cannot yield
any observable effect, as clearly shown in Eq.~(\ref{ampsq}).
It is to be emphasized that this feature is generic in any supersymmetric model
which ensures a light gravitino.
In the ordinary MSSM, this point is
hard to probe.
For example,
in the photino pair production,
double vertices of $e-\widetilde{e}-\phono$ in the
$t$- and $u$-channel Feynman diagrams eliminate the difference.

It is known that the availability of polarized electron and positron
beams is highly expected at future linear collider\cite{LC}:
The current LC performance goal is above 80\% of electron polarization
and 60\% of positron polarization.
We propose, therefore, that the effects of KK gravitinos
can be most sensitively measured by
\begin{equation}\label{Delsg}
\Delta \sg_{LR}
\equiv
\sg(e^-_L e^+_R \to \gm \rlap{\hspace{0.5mm}/}{E}_T)
-
\sg(e^-_R e^+_L \to \gm \rlap{\hspace{0.5mm}/}{E}_T)
\,.
\end{equation}
For the graviton production with a photon,
all the involved couplings are completely blind to the helicity
of the electron beam;
the $\Delta \sg_{LR}$ vanishes.
Moreover, in the SM,
the main contribution from the $Z$-pole to the
$\Delta \sg_{LR}$ is proportional to
$[(g_V+g_A)^2-(g_V-g_A)^2]=4g_V g_A$
where $g_V=-1/2+2\sin^2\theta_W$ and $g_A=-1/2$ \cite{Berends:1987zz}.
The smallness of $g_V$ suppresses the SM $Z$-pole background also.

\section{Numerical Results}
\label{Num}

The cross section of the process $e^+ e^-\to \phono\,\Gno$
obviously depends sensitively on the mass spectrum
of the involved supersymmetric particles,
a photino, and the left- and right-handed selectron.
Since the contribution of the $\widetilde{m}_{e_R}$
to the total cross section is very small,
two mass scales ($M_\phono$ and $\widetilde{m}_{e_L}$)
effectively determine the production rate.
One can obtain the mass spectrum of superparticles
by specifying a concrete supersymmetry breaking model,
such as the GMSB model which guarantees a light gravitino.
Instead we rather consider the experimental mass bounds
when the decay mode of $\widetilde{X} \to X \Gno$ is open:
At the LEP
the negative results of the $\gm \rlap{\hspace{0.5mm}/}{E}_T$
event search from  $e^+e^-
\to \Gno\,\widetilde{\chi}_1^0 \,\,(\widetilde{\chi}_1^0\to\gm\,\Gno)$
lead to $M_\phono \gsim 82.5$ GeV,
and those of the $\gm \gm \rlap{\hspace{0.5mm}/}{E}_T$
from  $e^+e^-\to \widetilde{\chi}_1^0\,
\widetilde{\chi}_1^0 \,\,(\widetilde{\chi}_1^0\to\gm\,\Gno)$
to $M_\phono \gsim 86.5$ GeV \cite{Groom:in}.
Similarly, the LEP bound with a light gravitino is
$\widetilde{m}_{e} \gsim 77$ GeV.
In the following numerical analysis,
we adopt the lower mass bounds of a photino and a left-handed selectron
as $M_\phono \geq 90$ GeV and $\widetilde{m}_{e_L} \geq 80$ GeV.
The $\widetilde{m}_{e_R}$ is set to be 200 GeV, which affects little
the total cross section.

\begin{figure}
\psfrag{XXXXLL2}[][][0.85]{{\large $M_\phono$ (GeV) }}
\psfrag{YYYYLL2}[][][0.85]{{\large $m_{\widetilde{e}_L}$ (GeV) }}
\psfrag{sigmin3}[][][0.85]{{ 0.1\,pb }}
\psfrag{sigmax3}[][][0.85]{{ 0.1\,fb }}
\psfrag{sigmin6}[][][0.85]{{ 1\,fb }}
\psfrag{sigmax6}[][][0.85]{{ 0.1\,fb }}
\psfrag{d3}[][][0.85]{{ \bf\Large (a) }}
\psfrag{d6}[][][0.85]{{ \bf\Large (b) }}
\includegraphics{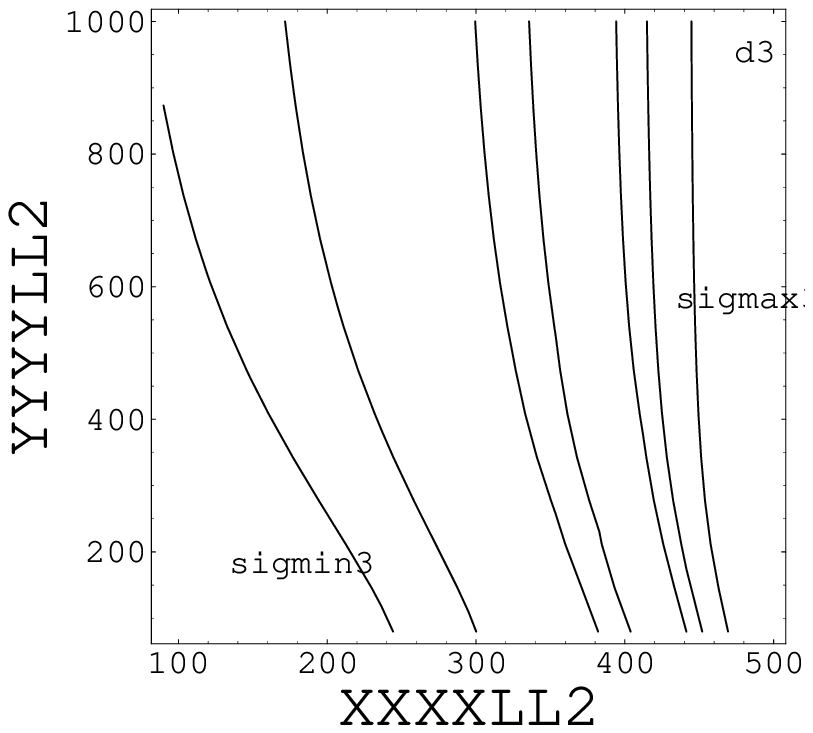}
\includegraphics{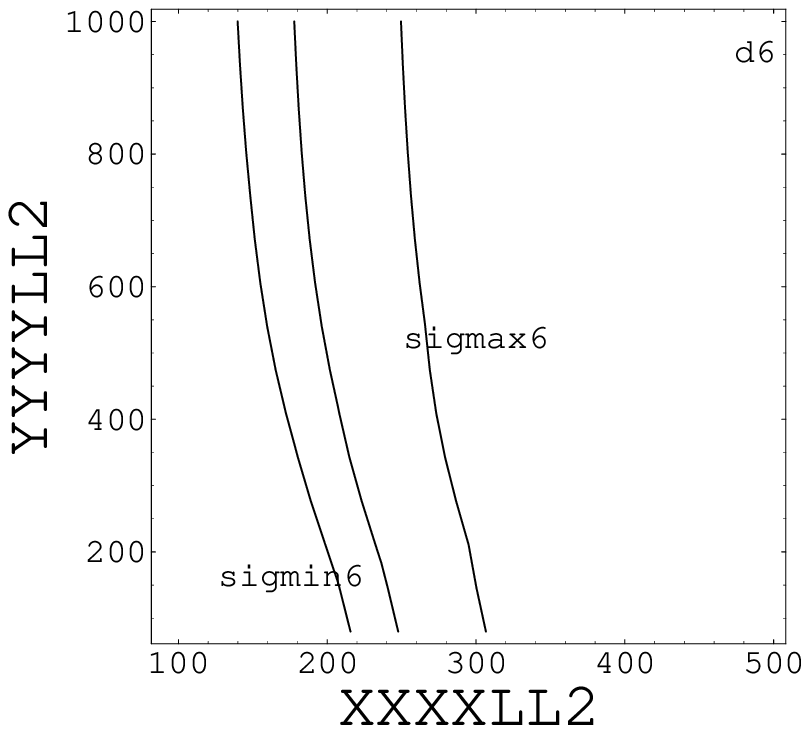}
\caption{\label{sg-Mphono-mse} At $\sqrt{s}=500$ GeV
with $M_S=1$ TeV,
the total cross section of $e^+ e^-\to\phono\Gno$
as a function of $M_\phono$ and $\widetilde{m}_{e_-}$.
Figure (a) is for $\dt=3$, with the contours from the left denoting
$\sg_{\rm tot}=100$, 50, 10, 5, 1, 0.5, 0.1 fb.
Figure (b) is for $\dt=6$
with $\sg_{\rm tot}=1, ~0.5,~0.1$ fb.}
\end{figure}

Figure \ref{sg-Mphono-mse}
presents the total cross section as a function of $M_\phono$ and
$\widetilde{m}_{e_L}$ for $\dt=3$ (a) and $\dt=6$ (b).
We set $\sqrt{s}=500$ GeV and $M_S=1$ TeV with the kinematic cut of
$|\cos\theta_\phono |<0.95$.
For the case of $M_\phono=90$ GeV and $\widetilde{m}_{e_L}=80$ GeV
where the total cross section reaches its maximum,
$\sg_{\rm tot}=321.4$ fb for $\dt=3$
and $\sg_{\rm tot}=8.8$ fb for $\dt=6$.
With the design luminosity of 500 (100) fb$^{-1}$/yr of the TESLA
(JLC and NLC)
\cite{Aguilar-Saavedra:2001rg,Abe:2001wn},
even the $\dt=6$ case with $M_\phono \lsim 300$
GeV can produce substantial events.
Being conservative,
we present the parameter space of
$(M_\phono,\widetilde{m}_{e_L})$ for $\sg_{\rm tot}>0.1$ fb.
It can been seen that in the $\widetilde{e}_L$ decoupling range with
$\widetilde{m}_{e_L} \gsim 500$ GeV,
the $s$-channel diagram alone can
produce sizable cross section.
As expected from the presence of light KK gravitinos,
this single photino production mode can probe the
photino mass much higher than $\sqrt{s}/2$,
the kinematic maximum for the photino pair production:
For $\dt=3$,
the photino with $M_\phono \lsim 460$ GeV
can be sufficiently produced;
for $\dt=6$, that with
$M_\phono \lsim 260$ GeV.
In the followings, we set $M_\phono=90$ GeV and $\widetilde{m}_{e_L}=80$ GeV.

\begin{figure}
\psfrag{XXXXLL3}[][][0.85]{{\large $\sqrt{s}$ (GeV) }}
\psfrag{YYYYLL3}[][][0.85]{{\large $\sigma$ (fb) }}
\psfrag{SML}[][][0.85]{{ $\sg^-_{_{SM}}$ }}
\psfrag{SMR}[][][0.85]{{ $\sg^+_{_{SM}}$ }}
\psfrag{GNL}[][][0.85]{{ $\sg^-(\Gno)$ }}
\psfrag{GNR}[][][0.85]{{ $\sg^+(\Gno)$ }}
\psfrag{G}[][][0.85]{{ $\sg^\mp(G)$ }}
\psfrag{FFF}[][][0.85]{{ $10^4$ }}
\psfrag{EEE}[][][0.85]{{ $10^3$ }}
\psfrag{DDD}[][][0.85]{{ $10^2$ }}
\psfrag{CCC}[][][0.85]{{ 10 }}
\psfrag{BBB}[][][0.85]{{ 1 }}
\psfrag{AAA}[][][0.85]{{ 0.1 }}
\psfrag{200}[][][0.85]{{ 200 }}
\psfrag{400}[][][0.85]{{ 400 }}
\psfrag{600}[][][0.85]{{ 600 }}
\psfrag{800}[][][0.85]{{ 800 }}
\psfrag{1}[][][0.85]{{ 1000 }}
\includegraphics{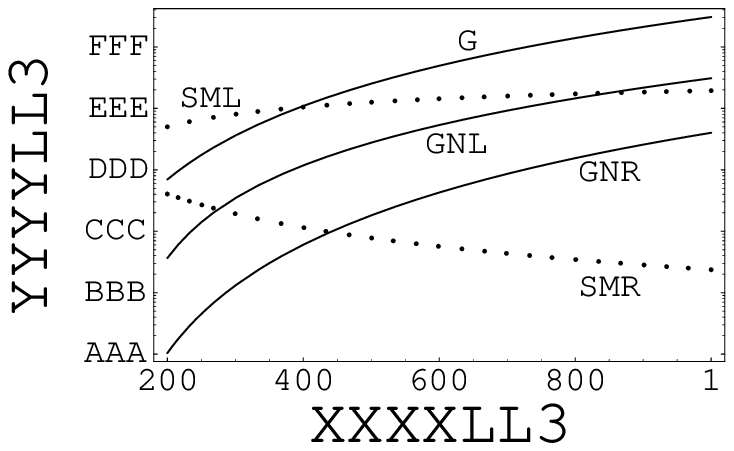}\caption{\label{fig3}
The polarized cross sections for the
$e^+ e^-\to\gm\nu\overline{\nu}$ in the SM denoted by $\sg_{SM}$,
$e^+ e^-\to\gm G^{\rm KK}$ by $\sg(G)$,
and $e^+ e^-\to\phono\Gno$ by $\sg(\Gno)$ when $\dt=3$.
The SM $\gm Z\to \gm\nu\overline{\nu}$ background
is reduced by a kinematic cut.
}
\end{figure}
\begin{figure}[thb]
\psfrag{XXXXLL6}[][][0.85]{{\large $\sqrt{s}$ (GeV) }}
\psfrag{YYYYLL6}[][][0.85]
   {{\large $\Delta\sigma_{_{LR}}^{ \widetilde{G}\widetilde{\gamma}}
   /\Delta\sigma_{_{LR}}^{ \rm SM}$ }}
\includegraphics{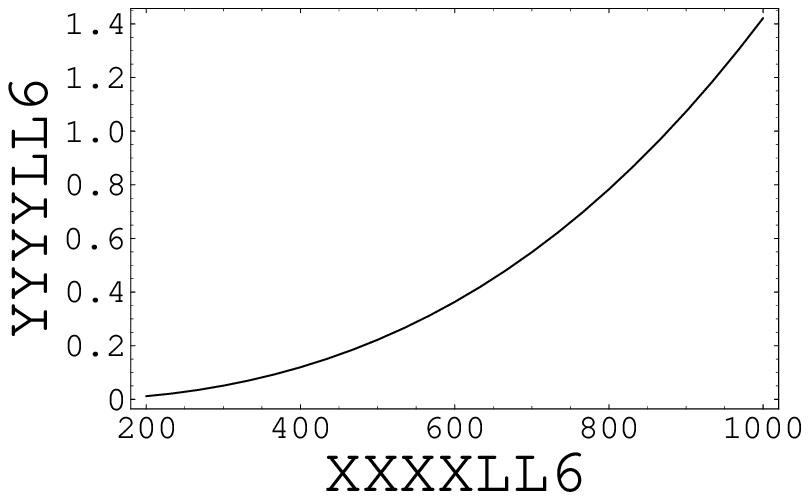}\caption{\label{fig-Delsg}
The ratio of the $\Delta \sg_{LR}(e^+ e^- \to \phono\Gno)$
to the $\Delta \sg_{LR}(SM)$ as a function of $\sqrt{s}$.}
\end{figure}

In Fig.~\ref{fig3}, we compare the polarized cross sections
of the single photon production at $e^+ e^-$ collisions,
with the neutrino pair in the SM (
$e^+ e^-\to\gm\nu\overline{\nu}$) denoted by $\sg_{SM}^\pm$,
with the KK gravitons ($e^+ e^-\to\gm G$) by $\sg^\pm(G)$,
and with the KK gravitinos ($e^+ e^-\to\phono\Gno$) by $\sg^\pm(\Gno)$.
Here superscript $\pm$ denotes the chirality of the electron beam.
To eliminate the SM $Z$-pole contribution as much as possible,
we employ the following kinematic cuts:
\begin{equation}\label{cuts}
20\,{\rm GeV}<E_{\gm(\phono)}< \frac{s-M_Z^2}{2\sqrt{s}}-20~{\rm GeV}\quad
{\rm and}\quad
|\cos\theta_{\gm (\phono)}|<0.95
\,.
\end{equation}
Since the $\sg_{SM}$ with the above cuts
are mainly through the $t$- and $u$-channel diagrams
mediated by the $W$ boson,
the $\sg_{SM}^+$
is much smaller than the $\sg_{SM}^-$.
For the KK graviton production,
the blindness of
the interactions of the graviton and the photon
to the fermion chirality guarantees the equality of
$\sg^-(G)$ and $\sg^+(G)$.
For the KK gravitino production,
there are several interesting points.
First its cross section is only a few tens percents of that for the KK graviton
production.
This is due to the kinematic suppression by the massive photino.
Second, the behavior of the cross section
with respect to $\sqrt{s}$ is the same as the KK graviton case,
which increases due to
the use of four-dimensional  \emph{effective} Lagrangian.
Finally the opposite sign of the photino coupling with the left- and right-handed
electron leads to the domination of the $\sg^-(\Gno)$ over the $\sg^+(\Gno)$.
In Fig.~\ref{fig-Delsg}, we present the ratio of
$\Delta \sg_{LR}(\phono\,\Gno)$ to
$\Delta \sg_{LR}(SM)$.
As discussed before,
the $\Delta \sg_{LR}$ vanishes for the KK graviton production.
Therefore, any deviation of the $\Delta \sg_{LR}$
from the SM background hints the presence of \emph{supersymmetric}
extra dimensions.
And this deviation increases with the beam energy.

\begin{figure}[thb]
\psfrag{XXXXLL4}[][][0.85]{{\large $x_{\widetilde{\gamma}}$ }}
\psfrag{YYYYLL4}[][][0.85]{{\large $d\sigma/d x_{\widetilde{\gamma}}$ (pb) }}
\psfrag{d3}[][][0.85]{{ $\dt=3$ }}
\psfrag{d4}[][][0.85]{{ $\dt=4$ }}
\psfrag{d5}[][][0.85]{{ $\dt=5$ }}
\includegraphics{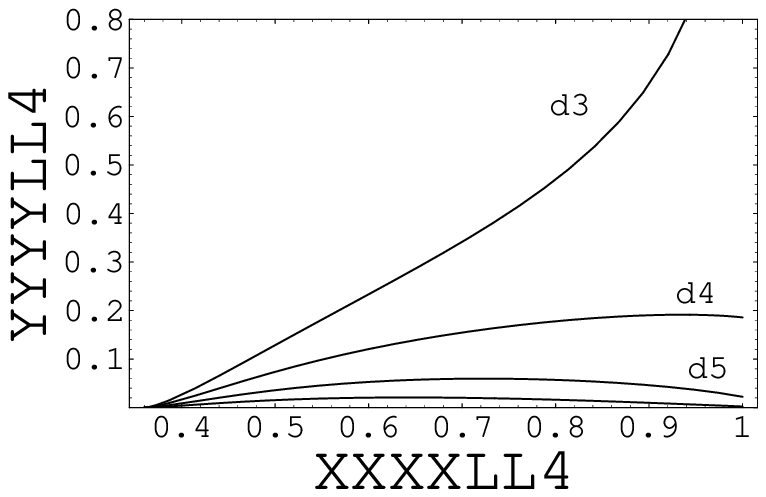}
\caption{\label{fig4} The differential cross section of the KK gravitino production
with respect to the photino energy fraction $x_\phono(\equiv 2 E_\phono/\sqrt{s})$.}
\end{figure}

Figure \ref{fig4} presents the
differential cross section of the KK gravitino production
with respect to the photino energy fraction
$x_\phono(\equiv 2 E_\phono/\sqrt{s})$ for various
$\dt$.
In the $\dt<4$ case,
a rapid increase occurs as the $x_\phono$ reaches its maximum;
energetic photinos are more likely produced.
This behavior can be understood from Eqs.~(\ref{sgtot}), (\ref{fGno}), and
(\ref{ampsq}).
Near the maximum of $x_\phono$,
light KK gravitinos are produced,
where the differential cross section behaves like
\begin{equation}
\label{limsg}
 \lim_{m^2\to 0}\frac{d \sg}{d x_\phono}
 \propto \lim_{m^2\to 0}
 \frac{(m^2)^{\dt/2-1}}{m^2}
,
\end{equation}
which the $m^2$ in the denominator
comes from the amplitude squared
in Eq.~(\ref{ampsq}).
The different behavior of the $\dt<4$ case is explained.
The measurement of this differential cross section
can tell whether the number of extra dimensions
is three or more.

\begin{figure}[thb]
\psfrag{XXXXLL5}[][][0.85]{{\large $z_\gamma$ }}
\psfrag{YYYYLL5}[][][0.85]{{\large $(1/\sg)d\sigma/d z_\gamma$ (pb) }}
\psfrag{SM}[][][0.85]{{ $\gm\nu\bar{\nu},\,\gm G$ }}
\psfrag{Gno}[][][0.85]{{ $\phono\Gno$ }}
\psfrag{SM6}[][][0.85]{{ $\gm\nu\bar{\nu},\,\gm G$ }}
\psfrag{Gno6}[][][0.85]{{ $\phono\Gno$ }}
\psfrag{RSa}[][][0.85]{{ $\sqrt{s}=200$ GeV }}
\psfrag{RSb}[][][0.85]{{ $\sqrt{s}=500$ GeV }}
\includegraphics{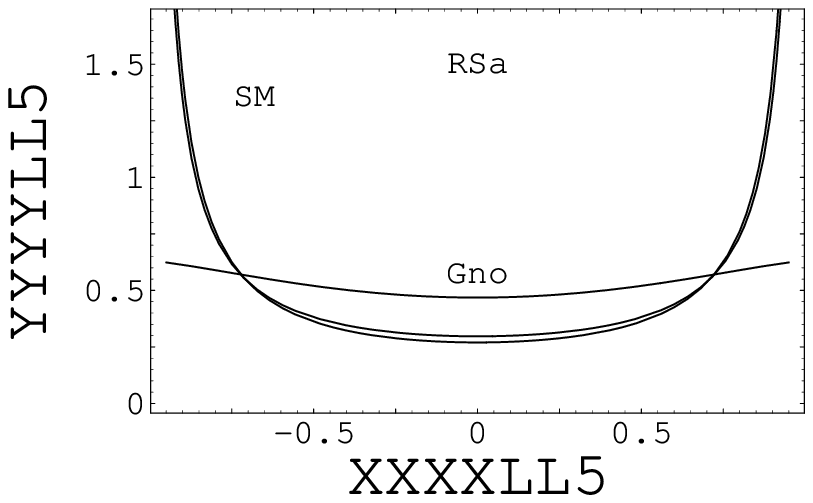}
\includegraphics{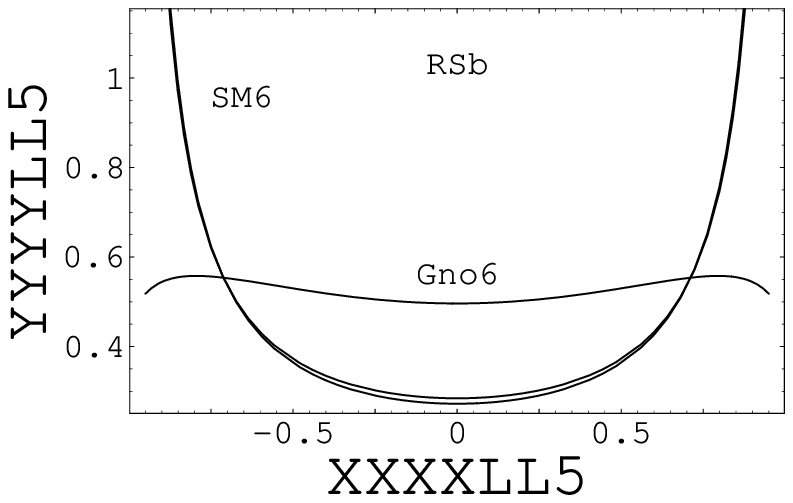}
\caption{\label{fig5} $(1/\sg)d \sg/d z_\gm$,
the angular distributions of the cross sections
for the SM, the KK graviton and KK gravitino production.
}
\end{figure}

In the $\dt=3$ case,
the scattering angle of the photino can be well approximated
by that of the photon decayed from the energetic photino.
Figure \ref{fig5} exhibits the angular distribution shapes for the $\dt=3$ case,
by plotting $(1/\sg)d \sg/d z_\gm$
with $z_\gm \equiv \cos\theta_\gm$.
The normalization by the total cross section
reveals the generic shape of the angular distribution.
For the SM and the KK graviton production,
the shapes are very similar:
Most of the photons are produced toward the beam line.
The KK gravitino production shows different behavior:
The angular distribution shape is rather flat.

\begin{table}
\caption{\label{table183} The $M_S$ bound in GeV from the $\sg_{\rm tot}$
with the kinematic cuts in Eq.~(\ref{cuts}) at $\sqrt{s}=183$ GeV and
the luminosity of 55.3 pb$^{-1}$ at 95\% CL. }
\begin{ruledtabular}
\begin{tabular}{ccccc}
 & $\dt=3$ &  $\dt=4$ &$\dt=5$ &$\dt=6$ \\
\hline
$G^{KK}$ & 764.5 & 621.5 & 525.6 & 457.5\\
$G^{KK}+\Gno^{KK}$ & 782.4 & 625.2 & 526.5 & 457.6 \\
\end{tabular}
\end{ruledtabular}
\end{table}

In Tables \ref{table183} and \ref{table500},
we summarize the sensitivity to the $M_S$ at 95\% CL
in two cases, when only the KK gravitons are produced and when
the KK gravitinos are also produced.
Table \ref{table183} is for
 $\sqrt{s}=183$ GeV with
the luminosity of 55.3 pb$^{-1}$,
and Table \ref{table500} for $\sqrt{s}=500$ GeV with
the luminosity of 100 fb$^{-1}$.
We have applied the kinematic cuts in Eq.~(\ref{cuts}).
With the KK gravitinos,
the increased cross section generally raises the sensitivity bound on the $M_S$.
Unfortunately,
the resulting change is practically negligible.

\begin{table}
\caption{\label{table500} The same $M_S$ bound in GeV at $\sqrt{s}=500$ GeV and
the luminosity of 100 fb$^{-1}$. }
\begin{ruledtabular}
\begin{tabular}{ccccc}
 & $\dt=3$ &  $\dt=4$ &$\dt=5$ &$\dt=6$ \\
\hline
$G^{KK}$ & 3250 & 2505 & 2037 & 1719\\
$G^{KK}+\Gno^{KK}$ & 3398 & 2559 & 2061 & 1732 \\
\end{tabular}
\end{ruledtabular}
\end{table}

\section{Conclusions}
\label{Conc}
Originally, extra dimensional models have been introduced
to solve the gauge hierarchy problem without resort to
supersymmetry.
However if the ultimate theory is
string theory, we live in
higher dimensional spacetime which has supersymmetry
as a fundamental symmetry.
And branes tend to break supersymmetry.
An interesting scenario is that
there are large and supersymmetric extra dimensions
and at least one supersymmetry survives on our brane below the scale
$M_S$ so that the low-energy effective theory on our brane
resembles
the MSSM.

The gravity supermultiplet resides in the bulk,
which includes the graviton and its super-partner, the gravitino.
On our brane, we have Kaluza-Klein towers of the graviton and gravitino.
If supersymmetry is not broken,
KK modes of the graviton would have the same mass spectrum
as those of the gravitino;
the zero mode of gravitino remains massless.
As the supersymmetry is broken by an expectation value of order $\Lmsusy$,
each gravitino KK mode acquires additional mass of $\Lmsusy^2/\Mpl$.
Under the assumption of low-energy $\Lmsusy$,
this mass shift
is sub-eV scale.
In practice, KK gravitinos
exist with almost continuous mass spectrum from zero.

In this scenario, we have studied the
KK gravitino production at $e^+ e^-$ collisions.
With $R$-parity conservation,
the KK gravitino is produced with a supersymmetric particle,
e.g., the photino.
Since light KK gravitinos become the LSP,
the photino decays into a photon and a KK gravitino (missing energy).
It has been shown that
the inclusive decay rate of $\phono\to\gm\,\Gno^\n$
is large enough for the photino
to decay within a detector.
Therefore,
the process $e^+ e^- \to \phono\,\Gno$ yields
a typical signature of a single photon with missing energy.
In the phenomenological allowed parameter space of
$(M_\phono,\widetilde{m}_{e_L})$,
we have shown that the total cross section can be substantial:
At $\sqrt{s}=500$ GeV,
$\sg_{\rm tot}>0.1$ fb for $M_\phono \lsim 460$ GeV in the $\dt=3$ case
and for $M_\phono \lsim 260$ GeV in the $\dt=6$ case.
The dependence of $\widetilde{m}_{e_L}$ is rather weak;
even in the range of
$\widetilde{m}_{e_L} \gsim 500$ GeV,
we have sizable cross section.

Unfortunately, the background processes
(the SM reaction of $e^+ e^- \to \gm \nu\overline{\nu}$ and
the KK graviton production of $e^+ e^- \to \gm G$) have much larger
cross sections.
With the $M_S$ of TeV,
the KK graviton production becomes compatible
with the SM background around $\sqrt{s}=500$ GeV.
However
the production of a massive photino kinematically suppresses
the KK gravitino production rate
compared to the KK graviton case by an order of magnitude,
since the $M_S$-dependence is the same.
To single out the effect of KK gravitinos,
total cross section is not enough.

We have noticed that the observable
$\Delta \sg_{LR} \equiv \sg(e^-_L e^+_R)-\sg(e^-_R e^+_L)$ can
completely eliminate the KK graviton background.
This is because both the gravitational and QED interactions,
which are involved in the KK graviton production,
do not distinguish the electron beam chirality;
$\Delta \sg_{LR} (\gm\,G)$ vanishes.
For the KK gravitino production accompanied by a photino,
the electron chirality becomes important since
the interaction of $e^-_L-\widetilde{e}_L-\phono$
has opposite sign to that of $e^-_R-\widetilde{e}_R-\phono$,
such that $\sg(e^-_L e^+_R) \gg \sg(e^-_L e^+_R)$.
The ratio of $\Delta \sg_{LR}(SM)$ to $\Delta \sg_{LR}(\phono\Gno)$
is demonstrated to increase with the beam energy,
implying that the observable $\Delta \sg_{LR}$ is unique and robust
to probe the \emph{supersymmetric} bulk.

We also found that the
differential cross section
with respect to the photino energy fraction $x_\phono$
can tell whether the number of extra dimensions
is three or more:
In the $\dt=3$ case,
the $d\sg/d x_\phono$ increases rapidly as $x_\phono$ approaches
its maximum;
energetic photinos are more likely produced.
And the angular distribution shapes, e.g., for the $\dt=3$ case,
are presented:
For the KK gravitino it is more or less flat, while
for the SM and the KK graviton
they rapidly increase toward the beam line.
The sensitivity bound of the $M_S$ at 95\% CL
does not practically change by taking into account of KK gravitino effects
due to the kinematic suppression of the
KK gravitino production cross section.

\appendix*
\section{The squared amplitudes of $e^+ e^-\to\phono\Gno$ }
For the process $e^+ e^-\to\phono\,\Gno^\n$,
the amplitudes squared in terms of the Mandelstam variables
defined in the text are
\begin{eqnarray}
\label{ampsq}
\left|
\widehat{\M}_s^\mp
\right|^2 &=&-  \frac{2}{3 s}\left( \frac{(t+u)(t^2+u^2)}{m^2_n} \right.
\\ \no && \left. \qquad
+2(s t +s u - 2 t u) +
           2 m_n \{m_n(m_n^2-M_\phono^2- s) +4 M_\phono s\}
           \right),
           \\ \no
\left|
\widehat{\M}_t^\mp
\right|^2 &=&\frac{2}{3m_n^2}
\frac{(M_\phono^2-t)(m_n^2-t)^3}{(\tilde{m}_{e_\mp}^2-t)^2},
\\ \no
\left|
\widehat{\M}_u^\mp
\right|^2 &=&\frac{2}{3m_n^2}
\frac{(M_\phono^2-u)(m_n^2-u)^3}{(\tilde{m}_{e_\mp}^2-u)^2},
\\ \no
2 \,\Re e\widehat{\M}_s^{\mp\dagger} \widehat{\M}_t^\mp
&=& \pm \frac{4}{3 m_n^2} \frac{1}{t-\tilde{m}_{e_\mp}^2}
\\ \no &\times&
\left[ t(m_n^2-t)^2
+m_n M_\phono
\{ m_n^2 (s-2t)+2t(s+t)+2M_\phono^2(m_n^2-t)\} \right],
\\ \no
2 \,\Re e\widehat{\M}_s^{\mp\dagger} \widehat{\M}_u^\mp
&=& \pm \frac{4}{3 m_n^2} \frac{1}{u-\tilde{m}_{e_\mp}^2}
\\ \no &\times&
\left[ u (m_n^2-u)^2
+m_n M_\phono
\{ m_n^2 (s-2u)+2u(s+u)+2M_\phono^2(m_n^2-u)\} \right],
\\ \no
2 \,\Re e\widehat{\M}_t^{\mp\dagger} \widehat{\M}_u^\mp
&=&-\frac{4 M_\phono s}{ 3 m_n (t-\tilde{m}_{e_\mp}^2) (u-\tilde{m}_{e_\mp}^2)  }
(2 M_\phono^2 m_n^2+m_n^2 s -2 t
u).
\end{eqnarray}

\acknowledgments
We thank Kingman Cheung, S.~Y.~Choi and E.~J.~Chun for
valuable discussions. The work of J.S. was
partially supported by the Korea Science and
Engineering Foundation (KOSEF) and the Deutsche Forschungsgemeinschaft
(DFG) through the KOSEF--DFG collaboration project, Project No.
20015--111--02--2

\def\MPL #1 #2 #3 {Mod. Phys. Lett. {\bf#1},\ #2 (#3)}
\def\NPB #1 #2 #3 {Nucl. Phys. {\bf#1},\ #2 (#3)}
\def\PLB #1 #2 #3 {Phys. Lett. {\bf#1},\ #2 (#3)}
\def\PR #1 #2 #3 {Phys. Rep. {\bf#1},\ #2 (#3)}
\def\PRD #1 #2 #3 {Phys. Rev. {\bf#1},\ #2 (#3)}
\def\PRL #1 #2 #3 {Phys. Rev. Lett. {\bf#1},\ #2 (#3)}
\def\RMP #1 #2 #3 {Rev. Mod. Phys. {\bf#1},\ #2 (#3)}
\def\NIM #1 #2 #3 {Nucl. Inst. Meth. {\bf#1},\ #2 (#3)}
\def\ZPC #1 #2 #3 {Z. Phys. {\bf#1},\ #2 (#3)}
\def\EJPC #1 #2 #3 {E. Phys. J. {\bf#1},\ #2 (#3)}
\def\IJMP #1 #2 #3 {Int. J. Mod. Phys. {\bf#1},\ #2 (#3)}
\def\JHEP #1 #2 #3 {J. High En. Phys. {\bf#1},\ #2 (#3)}


\begin{thebibliography}{99}
\bibitem{SUSY}
H.~E.~Haber and G.~L.~Kane,
Phys.\ Rept.\  {\bf 117} (1985) 75;
H.~P.~Nilles,
Phys.\ Rept.\  {\bf 110}, 1 (1984).


\bibitem{Antoniadis:1998ig}
N. Arkani-Hamed, S. Dimopoulos, and G. Dvali, \PLB B429 263 1998 ;
\PRD D59 086004 1999 ;
I. Antoniadis, N. Arkani-Hamed, S. Dimopoulos, and G. Dvali,  \PLB
B436 257 1998 .

\bibitem{Randall:1999ee}
L.~Randall and R.~Sundrum,
Phys.\ Rev.\ Lett.\  {\bf 83}, 3370 (1999).

\bibitem{Wells}
G.~F. Giudice, R. Rattazzi and J.~D. Wells, \NPB B544 3 1999 ;

\bibitem{ED-ph}
T.~Han, J.~D.~Lykken and R.~Zhang, Phys. Rev. {\bf D59}, 105006
(1999);
E.~A.~Mirabelli, M.~Perelstein and M.~E.~Peskin, Phys. Rev. Lett.
{\bf 82}, 2236 (1999);
J.~L.~Hewett, Phys. Rev. Lett. {\bf 82}, 4765 (1999);
K.~Y.~Lee, H.~S.~Song and J.~Song,
Phys.\ Lett.\ B {\bf 464}, 82 (1999);
K.~Y.~Lee, S.~C.~Park, H.~S.~Song, J.~Song and C.~Yu,
Phys.\ Rev.\ D {\bf 61}, 074005 (2000);
H.~Davoudiasl, J.~L.~Hewett and T.~G.~Rizzo,
Phys.\ Rev.\ Lett.\  {\bf 84}, 2080 (2000);
H.~Davoudiasl, J.~L.~Hewett and T.~G.~Rizzo,
Phys.\ Rev.\ D {\bf 63}, 075004 (2001).

\bibitem{susyed}
D.~Atwood, C.~P.~Burgess, E.~Filotas, F.~Leblond, D.~London and I.~Maksymyk,
Phys.\ Rev.\ D {\bf 63}, 025007 (2001);
D.~Marti and A.~Pomarol,
Phys.\ Rev.\ D {\bf 64}, 105025 (2001);
D.~E.~Kaplan and T.~M.~Tait,
JHEP {\bf 0006}, 020 (2000);
D.~E.~Kaplan, G.~D.~Kribs and M.~Schmaltz,
Phys.\ Rev.\ D {\bf 62}, 035010 (2000);
A.~Delgado, A.~Pomarol and M.~Quiros,
Phys.\ Rev.\ D {\bf 60}, 095008 (1999);
T.~Gherghetta and A.~Pomarol,
Nucl.\ Phys.\ B {\bf 586}, 141 (2000).


\bibitem{Altendorfer:2000rr}
R.~Altendorfer, J.~Bagger and D.~Nemeschansky,
Phys.\ Rev.\ D {\bf 63}, 125025 (2001);
J.~Bagger, D.~Nemeschansky and R.~J.~Zhang,
JHEP {\bf 0108}, 057 (2001);
T.~Gherghetta and A.~Pomarol,
Nucl.\ Phys.\ B {\bf 586}, 141 (2000), and
Nucl.\ Phys.\ B {\bf 602}, 3 (2001).




\bibitem{MTheory}
For reviews see, for example: J.~H.~Schwarz, Nucl.\
  Phys.\ Proc.\ Suppl.\ {\bf 55B}, 1 (1997) ;
  M.~J.~Duff, Int.\ J.\ Mod.\ Phys.\ {\bf A11}, 5623 (1996);
  J. Polchinski, hep-th/9611050; P.~K.~Townsend,
  hep-th/9612121; C. Bachas, hep-th/9806199; C.V. Johnson,
  hep-th/9812196.

\bibitem{arkani1}
N. Arkani-Hamed, L.J. Hall, Y. Nomura, D. Smith and
                   N. Weiner, hep-ph/0102090, {\it Nucl. Phys.} {\bf B605}
                   81 (2001);
E.~A.~Mirabelli and M.~E.~Peskin,
Phys.\ Rev.\ D {\bf 58}, 065002 (1998);
N.~Arkani-Hamed, T.~Gregoire and J.~Wacker,
JHEP {\bf 0203}, 055 (2002);
L.~Randall and R.~Sundrum,
Nucl.\ Phys.\ B {\bf 557}, 79 (1999);
Z.~Chacko, M.~A.~Luty, A.~E.~Nelson and E.~Ponton,
JHEP {\bf 0001}, 003 (2000);
A.~E.~Nelson and N.~J.~Weiner,
arXiv:hep-ph/0112210.

\bibitem{antoniadis}
J. Scherk and J.H. Schwarz, {\it Phys. Lett.} {\bf B82} 60
(1979); {\it Nucl. Phys.} {\bf B153} 61 (1979);
S. Ferrara, C. Kounnas, M. Porrati and F. Zwirner,
{\it Nucl. Phys.} {\bf B318} 75 (1989);
I.~Antoniadis,
Phys.\ Lett.\ B {\bf 246}, 377 (1990).
C. Kounnas and B. Rostand, {\it Nucl. Phys.} {\bf B341} 641 (1990);
E. Dudas and C. Grojean, hep-th/9704177, {\it Nucl. Phys.}
  {\bf B507} 553 (1997);
I. Antoniadis, S. Dimopoulos,
A. Pomarol and M. Quir\'os, {\it Nucl. Phys.} {\bf
B544} (1999) 503;
I.~Antoniadis, E.~Dudas and A.~Sagnotti,
Phys.\ Lett.\ B {\bf 464}, 38 (1999);
I.~Antoniadis, K.~Benakli and A.~Laugier,
arXiv:hep-th/0111209;
R.~Barbieri, L.~J.~Hall and Y.~Nomura,
Phys.\ Rev.\ D {\bf 63}, 105007 (2001).

\bibitem{Hewett:2002uq}
J.~L.~Hewett and D.~Sadri,
arXiv:hep-ph/0204063.

\bibitem{astro}
S.~Cullen and M.~Perelstein,
Phys.\ Rev.\ Lett.\  {\bf 83}, 268 (1999);
V.~D.~Barger, T.~Han, C.~Kao and R.~J.~Zhang,
Phys.\ Lett.\ B {\bf 461}, 34 (1999);
C.~Hanhart, D.~R.~Phillips, S.~Reddy and M.~J.~Savage,
Nucl.\ Phys.\ B {\bf 595}, 335 (2001);
L.~J.~Hall and D.~R.~Smith,
Phys.\ Rev.\ D {\bf 60}, 085008 (1999);
S.~Hannestad and G.~G.~Raffelt,
Phys.\ Rev.\ Lett.\  {\bf 88}, 071301 (2002), and
Phys.\ Rev.\ Lett.\  {\bf 88}, 071301 (2002).

\bibitem{Groom:in}
D.~E.~Groom {\it et al.}  [Particle Data Group Collaboration],
Eur.\ Phys.\ J.\ C {\bf 15}, 1 (2000).
\bibitem{Aguilar-Saavedra:2001rg}
J.~A.~Aguilar-Saavedra {\it et al.}  [ECFA/DESY LC Physics Working Group
                  Collaboration],
arXiv:hep-ph/0106315.
\bibitem{Abe:2001wn}
T.~Abe {\it et al.}  [American Linear Collider Working Group Collaboration],
in {\it Proc. of the APS/DPF/DPB Summer Study on the Future of Particle Physics (Snowmass 2001) } ed. R.~Davidson and C.~Quigg,
SLAC-R-570
{\it Resource book for Snowmass 2001, 30 Jun - 21 Jul 2001, Snowmass, Colorado}.

\bibitem{Mendez:1984qs}
A.~Mendez and F.~X.~Orteu,
Nucl.\ Phys.\ B {\bf 256}, 181 (1985);
P.~Fayet,
Phys.\ Lett.\ B {\bf 175}, 471 (1986);
P.~Fayet,
Phys.\ Lett.\ B {\bf 84}, 421 (1979).

\bibitem{GmGMSB}
G.~F.~Giudice and R.~Rattazzi,
Phys.\ Rept.\  {\bf 322}, 419 (1999);
S.~Ambrosanio, G.~L.~Kane, G.~D.~Kribs, S.~P.~Martin and S.~Mrenna,
Phys.\ Rev.\ D {\bf 54}, 5395 (1996);
D.~R.~Stump, M.~Wiest and C.~P.~Yuan,
Phys.\ Rev.\ D {\bf 54}, 1936 (1996).

\bibitem{LC}
J.~A.~Aguilar-Saavedra {\it et al.}  [ECFA/DESY LC Physics Working Group
                  Collaboration],
arXiv:hep-ph/0106315;
T.~Abe {\it et al.}  [American Linear Collider Working Group Collaboration],
in {\it Proc. of the APS/DPF/DPB Summer Study on the Future
of Particle Physics (Snowmass 2001) } ed. R.~Davidson and C.~Quigg,
arXiv:hep-ex/0106058.



\bibitem{Berends:1987zz}
F.~A.~Berends, G.~J.~Burgers, C.~Mana, M.~Martinez and W.~L.~van Neerven,
Nucl.\ Phys.\ B {\bf 301}, 583 (1988);
A.~D.~Dolgov, L.~B.~Okun and V.~I.~Zakharov,
Nucl.\ Phys.\ B {\bf 41} (1972) 197;
E.~Ma and J.~Okada,
Phys.\ Rev.\ Lett.\  {\bf 41}, 287 (1978)
[Erratum-ibid.\  {\bf 41}, 1759 (1978)];
K.~J.~Gaemers, R.~Gastmans and F.~M.~Renard,
Phys.\ Rev.\ D {\bf 19}, 1605 (1979).

\end{thebibliography}
\end{document}